\documentclass[cameraready]{Interspeech}
\title{Robust Wake-up Word Detection by Two-stage Multi-resolution Ensembles}

\author[affiliation={1, 2}, orcid=0000-0002-7705-2250, correspondingauthor]{Fernando}{López}
\author[affiliation={1}, orcid=0000-0002-4507-4930,]{Jordi}{Luque}
\author[affiliation={1}, orcid=0000-0001-6979-9330]{Carlos}{Segura}
\author[affiliation={1}]{Pablo}{Gómez}

\address{
    $^1$ Telefónica Innovación Digital, Spain \\
    $^2$ Universidad Autónoma de Madrid, Spain
}

\email{fernando.lopez@telefonica.com}

\keywords{Wake-up word detection, on-device, noise robustness, attention-based fusion}

\usepackage{comment,multirow}

\begin{document}

\maketitle

\begin{abstract}

    Voice interfaces depend on robust wake-up word (WuW) detection, yet noise robustness under tight on-device constraints remains challenging. We propose a two-stage detector from two complementary classifiers. The first stage is a lightweight streaming model that runs in real time, intended to minimize false negatives. When it fires, a verification stage runs to minimize false positives. The verification ensembles the hidden states of the triggering first-stage model with those of a stronger model via self-attention over time-preserving projections. The models differ in temporal resolution and architecture, yielding diverse embeddings; we hypothesize their fusion improves robustness. We select them by benchmarking accuracy and latency: the always-on stage runs in 25 ms and needs 144.4k operations, against 456.6M if the stronger model were run continuously. Under heavy noise, the ensemble matches or slightly improves on its stronger constituent, most clearly in the noisiest band.
    
\end{abstract}

\section{Introduction}
\label{sec:intro}
Voice-based interfaces have become a primary mode of hands-free interaction across consumer devices, and their usability depends on the WuW detector, which continuously listens for a trigger phrase and activates the system upon detection \cite{lopez2021deep}. This detector operates always-on and on-device; thus, it must satisfy strict latency, memory, and energy budgets while sustaining high detection accuracy \cite{macha2023fixed}. These requirements are difficult to meet in the presence of real-world acoustic degradation, including far-field capture, reverberation, speaker variability, and background noise. Such conditions arise at inference time and are only partially represented in the training data \cite{raju2018data, xiao2025ttakws}.

This work addresses noise robustness under tight on-device constraints, which remains a central challenge for WuW detection \cite{xi2025ntc}. Prior work improves robustness along three complementary axes. At the front-end, noise reduction \cite{cambara2022tase, lv2023dccrnkws, yang23robust}, contextual adaptation \cite{wang2017trainable}, and noise-robust learned features clean or stabilize the input before classification \cite{lopez2023filterbank}. During training, increased data and label diversity, targeted data augmentation, and loss functions tailored to noisy conditions expose the model to a broader range of acoustic variation \cite{raju2018data, xi2025ntc, dinkel2023unified, lopez2021novel}. At the model level, quantization and knowledge distillation render high-capacity architectures deployable \cite{macha2023fixed, Yang2023ondevice, mork2024noise}. In addition, cascade detection reduces cost and false alarms by pairing a lightweight on-device stage with a stronger verifier, deployed either on device or as a server-side two-pass network \cite{yashas2023to}; such verifiers are typically larger and more complex than the first-stage detector \cite{Kumar2020building}.

These approaches either optimize a single model or leverage the complementary strengths of architecturally distinct detectors only at the decision level. Although ensembling is a natural means of combining such strengths, prior WuW ensembles rely on spatial redundancy across multiple devices and fuse decisions by voting over predicted labels \cite{ahmed2022towards}. This strategy requires additional hardware and leaves the information contained in intermediate representations unexploited. We hypothesize that detectors differing in architecture and temporal resolution encode speech information differently, and that fusing their intermediate embeddings may therefore yield a representation more robust than any single constituent. Although attention-based fusion of intermediate embeddings has proven effective in other audio tasks, such as anti-spoofing \cite{rishith2025fusion}, the technique has not been studied under the constrained real-time conditions that characterize WuW detection.

Building on this hypothesis, we propose a two-stage detection scheme for noise robustness that targets real-time, computation-limited conditions. Our main contributions are as follows: (i) a WuW detector that pairs an always-on first stage with a verification stage, in which an attention-based network fuses the hidden states of heterogeneous detectors into time-preserving embeddings via self-attention, and the verifier reuses the hidden states of the first classifier; (ii) an empirical evaluation of the individual networks and their ensemble across a wide range of noise levels. (iii) a temporal-resolution study for the first stage that informs a favorable robustness-latency operating point; and (iv) a comparison of fourteen WuW models that characterizes their trade-off between real-time factor (RTF) and WuW F1 score, which informed our selection of the model combination in the ensemble.

\section{Two-stage multi-resolution ensemble}
The first stage is a lightweight classifier that processes the incoming audio stream continuously and in real time. We select for it the model with the best accuracy-latency trade-off. When it detects a candidate, it triggers the second stage, a verification network that confirms or rejects the detection.

This separation is central to our design. A single always-on detector must balance missed detections against false alarms at one fixed operating point, and a configuration light enough to run continuously typically forces a poor compromise. Decoupling triggering from verification instead exposes an independent operating point per stage. Because the second stage runs only on a trigger, its cost is confined to the moments that matter, leaving the always-on budget to fall on the first stage alone.

\begin{figure}[htp]
  \centering
  \centerline{\includegraphics[width=\columnwidth]{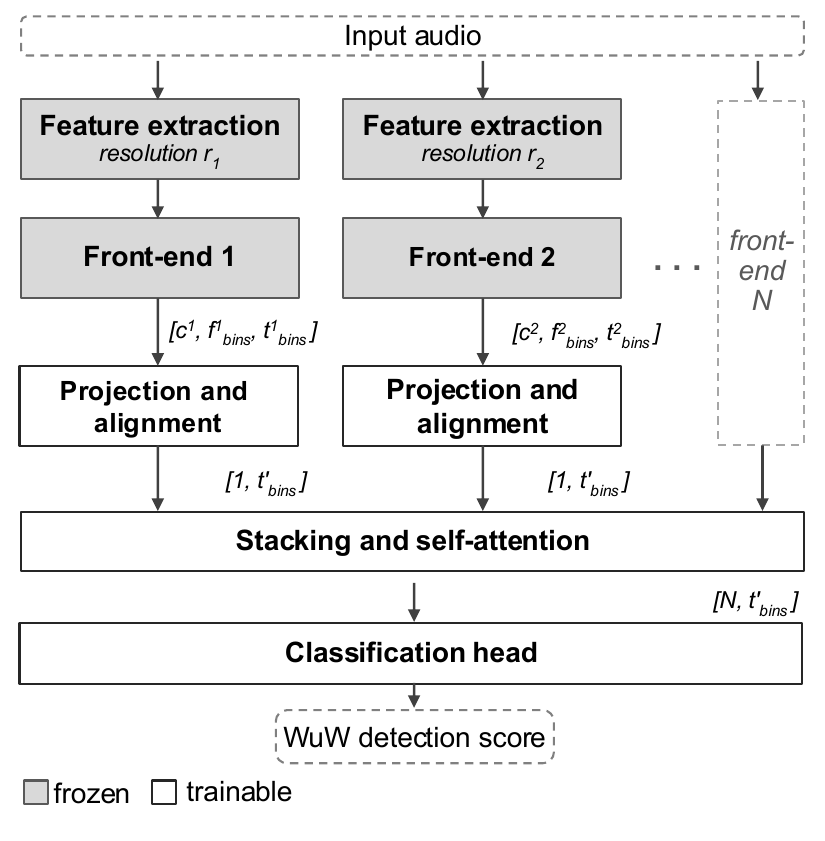}}
  \vspace{-0.45cm}
  \caption{Attention-based ensemble architecture. Each front-end is the frozen initial stack of a pre-trained classifier and extracts features at its own temporal resolution $r_i$. Projections align the embeddings to a common temporal length $t'_{bins}$, which self-attention then fuses. Front-end~1 is the always-on first-stage model, so its states are reused at no additional cost. Generalizes to $N$ front-ends, shown dashed; we use $N=2$.}
  \label{fig:ensemble-architecture}
\end{figure}


The verification stage reuses the intermediate embeddings that the first-stage classifier already computed for the triggering segment and fuses them with those of a second, heterogeneous classifier. Figure~\ref{fig:ensemble-architecture} shows the resulting network. Front-end~1 is the first-stage model, so its embeddings come at no additional cost, and verification adds only front-end~2 and the lightweight fusion network above them. The two front-ends differ in architecture and feature extraction, consuming features at different temporal
resolutions $r_1$ and $r_2$. Although the scheme generalizes to any number of front-ends $N$, as the dashed branch in Figure~\ref{fig:ensemble-architecture} indicates, each added classifier lengthens verification time under the real-time constraint; we therefore keep the count at two.

Prior work combines audio classifiers via voting \cite{ahmed2022towards}, Bayesian model averaging \cite{ristea2021self}, bagging
\cite{zahid2015optimized}, and stacking \cite{chen2022heart}. We adopt stacking, but instead of fusing classifier decisions, we fuse intermediate representations with attention, letting the verifier exploit each architecture's complementary strengths. Following Figure~\ref{fig:ensemble-architecture} from top to bottom, each front-end is the frozen initial stack of a pre-trained classifier, taken
before its final pooling and fully connected layers, and produces an embedding of shape $[c_i, f^i_{bins}, t^i_{bins}]$ for front-end $i$, where $c_i$ is the number of channels and $f^i_{bins}$ and $t^i_{bins}$ the frequency and temporal bins. Each projection and alignment block normalizes its embedding with a root-mean-square layer and maps the channel and frequency dimensions to a single value per frame through a fully connected layer with Swish activation, yielding $[1, t^i_{bins}]$ and preserving the temporal dimension. Because the
front-ends differ in temporal resolution, a 1D convolution then adapts each vector to $[1, t'_{bins}]$, the shortest temporal length across front-ends; the branch already at that length is left unchanged. Stacking the $N$ vectors yields a matrix of shape $[N, t'_{bins}]$, which a single-head self-attention layer with embedding dimension $d = N$ processes, weighing the front-ends against one another at each time frame. The classification head collapses the temporal axis by mean pooling to $[N, 1]$ and maps it to the detection score through a fully connected layer.

\section{Experiments}
\label{ssec:models}
We studied fourteen classifiers over thirteen architectures. Three baselines, \texttt{cnn-fat2019}, \texttt{cnn-trad-pool2}, and \texttt{sgru}, follow \cite{cambara2022tase}; the remaining ten span CNNs, RNNs, residual networks
(ResNets) \cite{tang2018deep}, Performers \cite{choromanski2020rethinking}, Broadcasted Residual Learning \cite{kim2021broadcasted}, Lambda Networks \cite{tura2021efficient}, and Conformers \cite{gulati2020conformer}.

\textbf{CNN and RNN-based models}\quad Alongside the convolutional baselines \texttt{cnn-fat2019} and \texttt{cnn-trad-pool2}, we studied Gated Recurrent Units (GRUs) under different post-processing of the temporal hidden states: the last memory state (\texttt{sgru}, the recurrent baseline), the maximum over time (\texttt{gru-max}), and a Performer applied to the states (\texttt{gru-att}).

\textbf{ResNets}\quad Residual networks with dilated convolutions: \texttt{resnet15}, \texttt{resnet15-narrow}, and \texttt{resnet8} \cite{tang2018deep}.

\textbf{Attention-based models}\quad \texttt{audiomer-l} \cite{sahu2021audiomer} processes raw audio directly; we replaced its attention kernels with the rectified linear activation function (ReLU). Building on the same work, we developed \texttt{audiomer-2d}, which processes Mel-Frequency Cepstral Coefficients (MFCC): a first convolutional block reduces the number of coefficients, followed by four Performer layers with convolutional residual connections. Convolutional layers produce the queries and context, and linear layers perform the classification. We further studied \texttt{lambda-resnet18}, which replaces the residual blocks of the ResNet with temporal Lambda layers; this front-end produces an embedding that a fully connected classification layer then processes.

\textbf{Broadcasted Residual Learning}\quad 
The \texttt{bc-resnet-1} network compresses and expands frequency information to perform 1D convolutions over time. We replace the average pooling of its final layer in~\cite{kim2021broadcasted} with a weighted average reduction.

\textbf{Conformers}\quad The \texttt{conformer} classifier has depth 12, 8 heads of size 64, a convolutional kernel size of 31, and a fully connected classification layer.

\begin{table}[t!]
\centering
\caption{\small Parameters, number of operations (multiplications and additions) and size of WuW detection models.}
\label{tab:wuw_classifiers}
\begin{tabular}{lccc}
\hline
\textbf{Model} & \textbf{Params.} & \textbf{Oper.}  &\textbf{Size (MB)}\\
\hline
cnn-fat2019	     & 5.2M    & 1196.8M 	& 43.6     \\ 
cnn-trad-pool2   & 69.5k   & 10.1M 		& 0.9      \\
\hline
gru-att	         & 643.9k  & 21.9M	    & 5.4     \\
gru-max 	     & 145.6k  & 21.5M	    & \textbf{0.8}     \\
sgru    	     & 145.6k  & \textbf{144.4k}	    & \textbf{0.8}     \\
\hline
resnet15	     & 237.4k  & 456.6M		& 10.6     \\
resnet15-narrow  & 42.4k   & 81.6M 		& 4.2      \\
resnet8	         & 109.8k  & 16.8M  	& 1.5      \\
\hline
bc-resnet-1	     & \textbf{9.4k}    & 4.0M 		& 5.7      \\
\hline
conformer        & 10.4M    & 107.4M      & 71.5     \\
audiomer-l	     & 674.2k  & 7.7M 		& 18.4     \\
audiomer-2d	     & 1.9M    & 3.3M 		& 10.9     \\
lambda-resnet18   & 89k     & 3.3M 		& 1.1        \\
\end{tabular}
\end{table}

Table \ref{tab:wuw_classifiers} reports all architectures by number of parameters, operations, and size. The \texttt{sgru} model has the fewest operations and the smallest size, which makes it a strong candidate for on-device execution.

\subsection{Database}
\label{ssec:databases}

The ``OK Aura'' database \cite{cambara2022tase} is used, retaining the test split as defined there and augmenting the training and validation splits with additional data from the M-AILABS Spanish database \cite{solak2019m-ailabs}, real room impulse responses (RIR) from SLR28 \cite{Ko2017reverberant}, and noise samples from \cite{valentini2016}. To extend the corpus, we recorded 260 in-house Spanish samples covering the target ``OK Aura'' key phrase and read speech, collected internally in compliance with the applicable data-protection regulations. The resulting data are split into training, validation, and test partitions following a 70\%/10\%/20\% ratio, with the original test split retained unchanged and no speaker overlap across splits. In total, the corpus comprises 70 hours of audio across $\sim$55k samples: the target keyword ($\sim$4k), other speech events ($\sim$26k), noise
($\sim$24k), and RIRs ($\sim$1k). A further 200 in-house samples, in which the trigger phrase is followed by a user utterance (e.g., ``OK Aura, quiero una pel\'icula estilo Star Wars''), were recorded and held out of this split; they are used exclusively to train and validate the attention-based fusion network.

\subsection{Audio preprocessing}
\label{ssec:audio-processing}
Every sample is normalized to a fixed 1.5-second window, chosen to cover the majority of the key-phrase duration distribution. For samples shorter than the window, the temporal position is randomized via zero-padding; for longer samples, a random 1.5-second segment is extracted. Peak normalization is then applied to ensure consistent amplitude levels. To simulate real-world conditions, every sample is contaminated with background noise, with the SNR drawn uniformly from the range $[-10, 50]$\,dB, and a fraction of samples is convolved with RIRs to reproduce diverse reverberant environments. As input features, we extract MFCCs rather than the Mel-spectrogram to obtain a more compact, decorrelated representation, replacing the zeroth coefficient with the log energy. Section \ref{ssec:feature-optimization} experiments with different numbers of coefficients and temporal resolutions to identify the best trade-off between robustness and inference time.

\subsection{Training and evaluation}
\label{ssec:training-and-evaluation}
All models share the fixed 1.5-second window for training and evaluation. We train from scratch, initializing the weights from a uniform distribution, for at most 700 epochs under a cross-entropy loss. Optimization uses Adam with a batch size of 128 and an initial learning rate of $10^{-3}$, reduced by one order of magnitude on a plateau; training stops after four consecutive learning-rate reductions without improvement.

For the attention-based ensemble, the two constituent models are selected from the test-split results of Section~\ref{sec:results}, and the front-ends of the pre-trained classifiers are kept frozen so that their hidden states can be reused directly. The batch size is reduced to 32 to accommodate the smaller pool of samples available for training the fusion network. The trainable parameters are therefore those of the projection, downsampling, self-attention, and classification.


We measure inference latency as the RTF on a Pixel XL device, the ratio of processing time to the duration of the processed audio. All models are compared at a common decision threshold of 0.5. We report classification performance using the WuW F1-score (the positive class). To characterize noise robustness, at test time we combine each test sample with held-out test noise and report it across eleven 5\,dB SNR bands spanning $[-10, 45]$\,dB.

\section{Results}
\label{sec:results}
\subsection{Feature optimization}
\label{ssec:feature-optimization}
\begin{figure}[t!]
  \centering
 \centerline{\includegraphics[width=0.9\columnwidth]{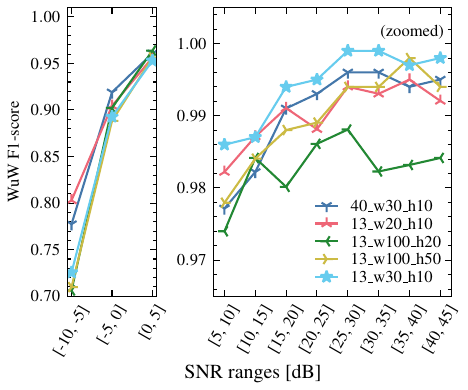}}
  \vspace{-0.35cm}
  \caption{Impact of temporal resolution and number of MFCCs on the WuW F1-score (positive class) by SNR range.}
\label{fig:data-processing}
\end{figure}

Feature-extraction parameters were selected for the first stage by measuring how the temporal resolution and the number of MFCC coefficients affect performance and inference cost. We trained and tested the \texttt{sgru} architecture with window sizes of 100\,ms, 30\,ms, and 20\,ms and hop sizes of 50\,ms, 20\,ms, and 10\,ms. Figure \ref{fig:data-processing} shows that the configurations are nearly indistinguishable at moderate-to-high SNR. Differences emerge almost exclusively in the noisiest ranges, where lower temporal resolution degrades the WuW F1-score most. Higher resolution, however, increases the amount of information to process and therefore the inference time. Table \ref{tab:feature-size} reports the feature size and the inference time on the Pixel XL for each configuration.

\begin{table}[b]
\centering
\caption{\small MFCC size and inference time on the Pixel XL Android device. The features column shows coefficients, window size (\textbf{w}) and hop size (\textbf{h}) in milliseconds.}
\label{tab:feature-size}
\begin{tabular}{lcc}
\hline
\textbf{Features} & \textbf{Size} & \textbf{Inference time (ms)} \\
\hline 
13, w=100, h=50  & \textbf{(29, 13)}    & \textbf{25.08}    \\ 
13, w=100, h=20	 & (71, 13)	   & 51.62    \\
13, w=30, h=10   & (148, 13)   & 73.43    \\
13, w=20, h=10   & (149, 13)   & 82.75        \\
40, w=30, h=10   & (148, 40)   & 83.97        \\
\end{tabular}
\end{table}

\begin{table*}[t!]
\caption{WuW F1-score (positive class) and the 95\% confidence intervals per SNR range for individual classifiers and the ensemble.}
\label{tab:ensemble}
\begin{minipage}[c]{\linewidth}
\centering
\begin{tabular}{ccccccc}
\hline
\multirow{2}{*}{
\parbox[c]{0.1\linewidth}{\centering SNR[dB]}}
  & \multicolumn{2}{c}{\textbf{device-sgru}} &
\multicolumn{2}{c}{\textbf{resnet15}} &
\multicolumn{2}{c}{\textbf{ensemble}} \\ 
\cmidrule(lr){2-3} \cmidrule(lr){4-5} \cmidrule(lr){6-7}
 & {\centering WuW f1-score} & {95\% CI} & {WuW f1-score} & {95\% CI}  & {WuW f1-score} & {95\% CI}\\
\midrule
{[}15, 20{]} & 0.988 & 0.977, 0.996 & \textbf{0.991} & 0.981, 0.998  & 0.989 & 0.976, 0.996 \\
{[}10, 15{]} & 0.984 & 0.969, 0.994 & \textbf{0.993} & 0.984, 0.999 & 0.988 & 0.979, 0.995 \\
{[}5, 10{]} & 0.978 & 0.961, 0.990 & 0.991 & 0.978, 0.999 & \textbf{0.993} & 0.989, 0.997 \\
{[}0, 5{]} & 0.956 & 0.934, 0.973 & 0.982 & 0.967, 0.992 & \textbf{0.983} & 0.972, 0.989 \\
{[}-5, 0{]} & 0.888 & 0.841, 0.923 & 0.944 & 0.918, 0.965 & \textbf{0.946} & 0.929, 0.968 \\
{[}-10, -5{]} & 0.710 & 0.627, 0.779 & 0.829 & 0.764, 0.879 & \textbf{0.843} & 0.812, 0.898 \\
\midrule
{Full range} & 0.917 & 0.885, 0.943 & 0.955 & 0.932, 0.972 & \textbf{0.957} & 0.934, 0.974 \\
\end{tabular}
\end{minipage}
\end{table*}

Given this trade-off, we selected 13 MFCCs with a 100\,ms window and a 50\,ms hop for the first stage. This configuration sacrifices little accuracy above 0 dB, while keeping inference at $\sim$25\,ms, which suits always-on execution. We refer to the \texttt{sgru} trained with this configuration as \texttt{device-sgru}. For the verification network, we chose a deliberately different configuration: 40 MFCCs with a 30\,ms window and a 10\,ms hop, which gave the best results. This difference is intentional: combined with the distinct architectures of the two stages, they operate at different temporal resolutions and encode speech differently, providing the representational diversity. Given that the verifier runs only on a trigger, its feature cost does not affect the always-on budget.

\subsection{Individual classifiers and ensemble}
\label{ssec:ensemble}
\begin{figure}[t!]
  \centering
  \centerline{\includegraphics[width=7.2cm, height=3.9cm]{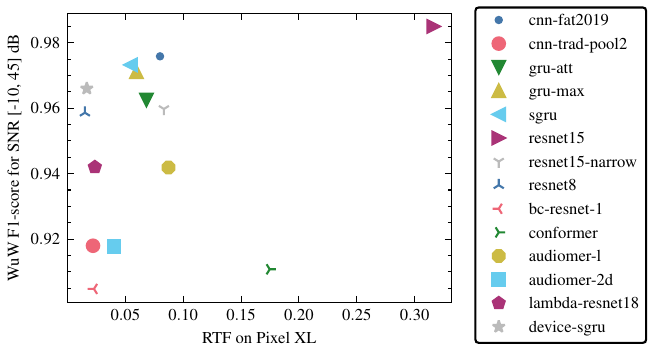}}
  \vspace{-0.27cm}
\caption{WuW F1-score (positive class) over the $[-10, 45]$ dB SNR range vs.\ RTF on the Pixel XL.}
\label{fig:performance-vs-rtf}
\end{figure}

\begin{figure}[h]
  \centering
  \centerline{\includegraphics[width=6.0cm, height=5.2cm]{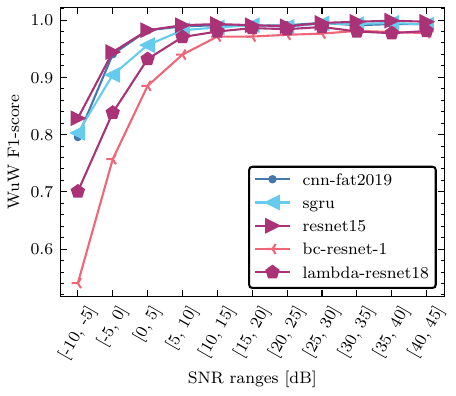}}
  \vspace{-0.4cm}
\caption{WuW F1-score (positive class) per SNR range for the best-performing classifiers.}
\label{fig:resistance-against-noise}
\end{figure}

Figure \ref{fig:performance-vs-rtf} compares all classifiers by RTF and WuW F1-score. We trained them on the same features (40 MFCCs, a 30\,ms window, and a 10\,ms hop), except for \texttt{audiomer-l}, which processes raw audio, and \texttt{device-sgru}, described in Section \ref{ssec:feature-optimization}. The figure confirms that \texttt{device-sgru} offers the best trade-off, combining a high F1-score with one of the smallest RTFs. \texttt{resnet15} attains the highest F1-score of any individual classifier. We select it as the stronger, heterogeneous model to pair with \texttt{device-sgru} in the verification stage. Figure \ref{fig:resistance-against-noise} shows the best-performing classifiers under noise: the gap between models is widest in the noisiest SNR ranges and narrows as the noise weakens.

The ensemble fuses \texttt{device-sgru} with \texttt{resnet15}, adding only $\sim$6.2k trainable parameters. Table \ref{tab:ensemble} reports the results. The two constituents perform comparably at high SNR, but their behavior diverges as noise increases. From $[5,10]$\,dB downward the ensemble leads, and the margin is largest in the harshest range, $[-10,-5]$\,dB, where \texttt{device-sgru} falls to 0.710 and \texttt{resnet15} to 0.829, while the ensemble reaches 0.843, above its stronger constituent, though the intervals overlap. This is consistent with our hypothesis that fusing diverse embeddings can aid robustness where a single detector is most fragile. Across the full SNR range, the ensemble and \texttt{resnet15} are effectively tied, with both clearly above \texttt{device-sgru}. This parity is expected, since the two diverge only in the lowest bands; the full-range figure therefore understates the effect that matters for robustness.
In terms of latency, the always-on first stage runs in $\sim$25\,ms and adds no perceptible delay. When it fires, extracting the verifier's features takes $\sim$13\,ms and the verification network runs in $\sim$475\,ms. The verification is triggered only on positives and overlaps the user's continuing utterance. Thus, this cost does not delay genuine activations on device. The same cascade also admits a device-server configuration, where the lightweight stage runs on device and the verifier is offloaded.

\section{Limitations}
The ensemble was trained on a small reserved set; since most of its trainable parameters lie in the projection mapping each embedding to its temporal vector, more data should yield richer representations. Although the ensemble's point estimate leads its strongest constituent in noisier bands, the per-model confidence intervals overlap, so we report a consistent trend rather than a significant per-band improvement. We fused a single pair, \texttt{device-sgru} and \texttt{resnet15}, and did not compare against a cascade in which \texttt{resnet15} verifies alone. Which heterogeneous pairs fuse best, whether more than two constituents help, and setting a per-stage operating point in a streaming cascade are future work. 

\section{Conclusions}
\label{sec:conclusions}
We presented a noise-robust, two-stage scheme for wake-up word detection that runs on device. The always-on first stage runs in 25 ms and needs 144.4k operations, against 456.6M if the stronger model were run continuously, and the second stage is an attention-based verification network that fuses the intermediate embeddings of heterogeneous classifiers computed at different temporal resolutions. The fusion aligns the embeddings to a common temporal resolution and processes the stacked vectors with self-attention. The ensemble matches, and in the harshest bands slightly leads, the strongest individual classifier under noise, suggesting that embedding diversity is a promising source of robustness.

\section{Generative AI Use Disclosure}
We used a generative AI to assist in paraphrasing and improving clarity and grammar in parts of the manuscript; all generated content was carefully reviewed and validated by the authors.



\end{document}